\documentstyle[a4,12pt]{article}
\renewcommand{\baselinestretch}{1.3}

\def\be{\begin{equation}}
\def\ee{\end{equation}}
\def\beq{\begin{eqnarray}}
\def\eeq{\end{eqnarray}}
\def\dis{\displaystyle}
\def\l{\left}
\def\r{\right}
\def\ba{\begin{array}}
\def\ea{\end{array}}
\def\rw{\rightarrow}
\def\la{\langle}
\def\ra{\rangle}
\def\f{\frac}
\def\s{\sqrt}

\def\U{\Upsilon (4S)}
\def\Pa{P^{0}}
\def\Pb{\bar{P}^{0}}
\def\Pap{\Pa_{\rm phys}}
\def\Pbp{\Pb_{\rm phys}}
\def\pp{\Pap\Pbp}
\def\Ba{B^{0}_{d}}
\def\Bb{{\bar B}^{0}_{d}}
\def\Bap{B^{0}_{d, {\rm phys}}}
\def\Bbp{{\bar B}^{0}_{d, {\rm phys}}}
\def\pair{\Bap\Bbp}
\def\a{\alpha}
\def\b{\beta}
\def\G{\Gamma}
\def\p{\phi}
\def\lm{\lambda_{f}}
\def\e{{\rm exp}}
\def\D{\Delta}
\def\z{\zeta}
\def\x{\xi}
\def\Xc{X_{\bar{c}c}}

\begin{document}

\begin{flushright}
{\large\bf LMU-22/94}\\
{December 1994}
\end{flushright}

\begin{center}
{\Large\bf Time Dependence of Coherent $\; P^{0}\bar{P}^{0} \;$ Decays \\
and $CP$ Violation at Asymmetric $B$ Factories}
\end{center}

\vspace{2cm}

\begin{center}
{\bf Zhi-zhong XING}$^{~*}$
\end{center}

\begin{center}
{\sl Sektion Physik, Theoretische Physik, Universit${\sl\ddot a}$t M${\sl\ddot
u}$nchen}\\
{\sl Theresienstrasse 37, D-80333 Munich, Germany}
\end{center}

\vspace{2.6cm}

\begin{abstract}

A generic formalism is presented for the time-dependent or time-integrated
decays of any coherent $P^{0}\bar{P}^{0}$ system ($P^{0}= K^{0}, D^{0},
B^{0}_{d}$, or $B^{0}_{s}$).
To meet various possible measurements at asymmetric $B$ factories,
we reanalyze some typical signals of $CP$ violation in the coherent
$B^{0}_{d}\bar{B}^{0}_{d}$
transitions. The advantage of proper time cuts is illustrated
for measuring mixing parameters and $CP$ violation. We show that the
direct and indirect $CP$ asymmetries are distinguishable in neutral $B$ decays
to $CP$ eigenstates.
The possibility to detect the $CP$-forbidden processes at the $\U$ resonance is
explored in
some detail.

\end{abstract}

\vspace{4.6cm}

\begin{flushleft}
{-------------------------------------------------}\\
{$^{*}$ {\footnotesize Alexander von Humboldt Research Fellow}}\\
{$^{*}$ {\footnotesize E-mail: Xing$@$hep.physik.uni-muenchen.de}}
\end{flushleft}

\newpage

\begin{flushleft}
{\Large\bf 1. Introduction}
\end{flushleft}

It is known in particle physics that mixing between a neutral meson
$\Pa$ and its $CP$-conjugate state $\Pb$ provides a mechanism whereby
interference
in the decay amplitudes can occur, leading to the possibility of $CP$ violation
[1].
To date, $K^{0}-\bar{K}^{0}$ mixing and $B^{0}_{d}-\bar{B}^{0}_{d}$ mixing have
been measured,
and the $CP$ violating signal induced by $K^{0}-\bar{K}^{0}$ mixing
has been unambiguously established [2]. Compared with
$B^{0}_{d}-\bar{B}^{0}_{d}$ mixing,
$B^{0}_{s}-\bar{B}^{0}_{s}$ ($D^{0}-\bar{D}^{0}$) mixing is expected to be
quite large
(very small) in the context of the standard electroweak model [3].
Today the $\Ba-\Bb$ system is playing an important role in studying flavour
mixing and
$CP$ violation beyond the neutral kaon system. The $B^{0}_{s}-\bar{B}^{0}_{s}$
and $D^{0}-\bar{D}^{0}$ systems are more interesting in practice for probing
new physics
that is out of reach of the standard model predictions.

A feasible way to study $CP$ violation arising from $\Pa-\Pb$ mixing is to
measure
the coherent decays of $\Pa\Pb$ pairs produced at appropriate resonances. For
instance,
$$
\phi \; \rw \; K^{0}\bar{K}^{0} \; , \;\;\;\;\;\; \psi^{''} \; \rw \;
D^{0}\bar{D}^{0}
\; , \;\;\;\;\;\; \U \; \rw \; \Ba\Bb \; , \;\;\;\;\;\;
\Upsilon (5S) \; \rw \; B^{0}_{s}\bar{B}^{0}_{s} \; .
$$
At present, efforts are underway to develop asymmetric $B$ factories at
KEK and SLAC laboratories [4], while asymmetric $\phi$ factory options are also
under consideration [5].
The main purpose of these machines is to probe $CP$ violation (and to test
other discrete
symmetries or conservation laws) by measuring the time-dependent
transitions. By now some phenomenological analyses of
coherent $K^{0}\bar{K}^{0}$ and $\Ba\Bb$ decays have been made in the
literature [6-8]. These works
have outlined the main features of $CP$ violation in the $K^{0}-\bar{K}^{0}$ or
$\Ba-\Bb$ systems, although many of their formulae and results rely on the
characteristics of the system itself or some model-dependent approximations.
A generic formalism, which can describe the common properties
of coherent $\Pa\Pb$ decays, is still lacking. In addition, little attention
has been paid to
the advantage of proper time cuts for measuring the mixing parameters and $CP$
asymmetries in coherent weak decays.

In this paper we shall present a generic and concise formalism for the
time-dependent or time-integrated
decays of any coherent $\Pa-\Pb$ system.
This formalism should be very useful for phenomenological applications, because
it
is independent of the contexts of the standard model or its various
non-standard
extensions. To meet various possible measurements at the forthcoming
$B$ factories, we shall carry out a reanalysis of the typical signals of $CP$
violation manifesting in
$\Ba-\Bb$ mixing, in $\Ba$ vs $\Bb$ decays to $CP$ eigenstates, and in
$CP$-forbidden
transitions at the $\U$ resonance. Our analysis differs from the previous ones
in the following
three aspects: (1) we illustrate the advantage of proper time cuts
for measurements of the mixing parameters and $CP$ violating asymmetries;
(2) we highlight the distinguishable effect of direct $CP$ violation on $CP$
asymmetries
in neutral $B$ decays; and (3) we explore in some detail the possibility to
detect the $CP$-forbidden decays
at $B$ factories. This work concentrates on analytical studies.
A more comprehensive discussion about the present topic, together with
numerical (model-dependent) predictions,
will be given elsewhere. \\

\begin{flushleft}
{\Large\bf 2. Formalism of coherent $\Pa\Pb$ decays}
\end{flushleft}

	The time-dependent wave function for a $\pp$ pair at rest can be written as
\begin{equation}
\frac{1}{\sqrt{2}}\left [ |\Pap({\bf K},t)\rangle \otimes |\Pbp(-{\bf
K},t)\rangle
+C |\Pap(-{\bf K},t)\rangle \otimes |\Pbp({\bf K},t)\rangle \right ] \; ,
\end{equation}
where $\bf K$ is the three-momentum vector of the $P$ meson, and $C=-$ or $+$
is the
charge-conjugation parity of the $\pp$ pair.
The proper time evolution of an initially ($t=0$) pure $\Pa$ or $\Pb$ is given
by
\be
\ba{rcl}
|\Pap(t)\ra & = & g_{+}(t)|\Pa\ra + (q/p) g_{-}(t)|\Pb\ra \; , \\
|\Pbp(t)\ra & = & (p/q) g_{-}(t)|\Pa\ra + g_{+}(t)|\Pb\ra \; ,
\ea
\ee
where the mixing parameters $p$ and $q$ connect the flavour eigenstates
$|\stackrel{(-)}{P}$$^{0}\ra$ to the mass eigenstates $|P_{1,2}\ra$
through $|P_{1}\ra = p|P^{0}\ra + q|\bar{P}^{0}\ra$ and
$|P_{2}\ra  = p|P^{0}\ra - q|\bar{P}^{0}\ra$; and
\be
g_{\pm}(t) \; = \; \f{1}{2} e^{- \l (im +\f{\G}{2}\r )t}
\l [ e^{+ \l (i\D m -\f{\D \G}{2}\r ) \f{t}{2}}
\pm e^{- \l (i\D m -\f{\D \G}{2}\r ) \f{t}{2}} \r ]  \; .
\ee
Here we have defined $m = (m_{1}+m_{2})/2$, $\D m  = (m_{2}-m_{1})$,
$\G = (\G_{1}+\G_{2})/2$, and $\D \G  = (\G_{1}-\G_{2})$,
where $\G_{1,2}$ and $m_{1,2}$ are the widths and masses of $P_{1,2}$ .

	Now we consider the case that one of the two $P$ mesons
(with the momentum $\bf K$) decays to a final state $f_{1}$ at proper time
$t_{1}$ and the other (with $-\bf K$) to $f_{2}$ at $t_{2}$ .
$f_{1}$ and $f_{2}$ may be either hadronic or semileptonic states.
After a lengthy calculation, the joint decay rate for having such an event is
given as
\be
\ba{rcl}
{\rm R}(f_{1}, t_{1}; f_{2}, t_{2})_{C} & \propto &
|A_{f_{1}}|^{2} |A_{f_{2}}|^{2} e^{-\G t_{+}}
\dis\left [ \f{1}{2} |\x_{C}+\z_{C}|^{2} e^{-\f{\D \G}{2}t_{C}}
+ \f{1}{2} |\x_{C}-\z_{C}|^{2} e^{+\f{\D \G}{2}t_{C}}  \r . \\
&  & \l . - \dis\l (|\x_{C}|^{2}-|\z_{C}|^{2}\r ) \cos (\D mt_{C})
+ 2{\rm Im}\l (\x_{C}^{*}\z_{C}\r ) \sin (\D mt_{C}) \r ] \; ,
\ea
\ee
where
\be
A_{f_{i}} \; = \; \la f_{i}|H|\Pa \ra \; , \;\;\;\;\;
\bar{A}_{f_{i}} \; =\; \la f_{i}|H|\Pb \ra \; , \;\;\;\;\;
\rho^{~}_{f_{i}} \; =\; \f{\bar{A}_{f_{i}}}{A_{f_{i}}} \; , \;\;\;\;\;
(i=1, 2) \; ;
\ee
and
\be
t_{C} \; =\; t_{2}+Ct_{1} \; ,\;\;\;\;\;
\x_{C} \; =\; (p/q) +C (q/p) \rho^{~}_{f_{1}}\rho^{~}_{f_{2}} \;
\;\;\;\;\; \z_{C} \; =\; \rho^{~}_{f_{2}}+C \rho^{~}_{f_{1}} \; .
\ee
The time-independent decay rate is obtainable from Eq. (4) by integrating
${\rm R}(f_{1},t_{1};f_{2},t_{2})_{C}$ over $t_{1}$ and $t_{2}$ :
\be
\ba{rcl}
{\rm R}(f_{1}, f_{2})_{C} & \propto & |A_{f_{1}}|^{2}|A_{f_{2}}|^{2} \l [
\dis\f{ |\x_{C}+\z_{C}|^{2} }{2(1+y)(1+Cy)} +
\dis\f{ |\x_{C}-\z_{C}|^{2} }{2(1-y)(1-Cy)} \r . \\
& & \l . -\dis\f{1-Cx^{2}}{(1+x^{2})^{2}} \l ( |\x_{C}|^{2} - |\z_{C}|^{2} \r )
+ \dis\f{2(1+C)x}{(1+x^{2})^{2}} {\rm Im}\l (\x^{*}_{C}\z_{C} \r ) \r ] \; ,
\ea
\ee
where $x = \D m/\G$ and $y = \D \G/(2\G)$ are two measurables of the
$\Pa-\Pb$ system. Note that Eqs. (4) and (7) are useful at both symmetric and
asymmetric flavour factories.

	An asymmetric $e^{+}e^{-}$ collider running at the threshold of production of
$(\pp)_{C}$ pairs
will offer the possibility to measure the decay-time difference
$t_{-}=(t_{2}-t_{1})$
between $\Pap\rw f_{1}$ and $\Pbp\rw f_{2}$ . It is usually difficult to
measure the
$t_{+}=(t_{2}+t_{1})$ distribution in either linacs or storage rings,
unless the bunch lengths are much shorter than the decay
lengths [4,5,9]. Hence it is more practical to study the $t_{-}$ distribution
of the joint decay rates.
Here and hereafter we use $t$ to denote $t_{-}$ for simplicity.
Integrating ${\rm R}(f_{1},t_{1};f_{2},t_{2})_{C}$ over $t_{+}$, we obtain
the decay rates (for $C=\pm $) as follows:
\be
\ba{rcl}
{\rm R}(f_{1},f_{2}; t)_{-} & \propto & |A_{f_{1}}|^{2}|A_{f_{2}}|^{2}
e^{-\G |t|}\dis\l [ \f{1}{2} |\x_{-}+\z_{-}|^{2} e^{-y \G t}
+ \f{1}{2} |\x_{-}-\z_{-}|^{2} e^{+y \G t} \r . \\
&  & \l .  - \dis\l (|\x_{-}|^{2}-|\z_{-}|^{2}\r ) \cos (x\G t)
+ 2{\rm Im}\l (\x_{-}^{*}\z_{-}\r ) \sin (x\G t) \r ] \;
\ea
\ee
and
\be
\ba{rcl}
{\rm R}(f_{1},f_{2}; t)_{+} & \propto & |A_{f_{1}}|^{2}|A_{f_{2}}|^{2}
e^{-\G |t|} \l [ \dis\f{ |\x_{+}+\z_{+}|^{2} }{2(1+y)} e^{-y\G |t|}
+ \dis\f{ |\x_{+}-\z_{+}|^{2} }{2(1-y)} e^{+y\G |t|} \r . \\
& & \l .  - \dis\f{ |\x_{+}|^{2}-|\z_{+}|^{2} }{\sqrt{1+x^{2}}}
\cos \l (x\G |t| +\phi_{x}\r ) + \dis\f{ 2{\rm Im}\l (\x^{*}_{+}\z_{+}\r ) }
{\sqrt{1+x^{2}}} \sin (x\G |t| +\phi_{x}) \r ]  \; ,
\ea
\ee
where $\phi_{x}= \arctan x$ signifies a phase shift. One can
check that integrating ${\rm R}(f_{1},f_{2};t)_{C}$ over $t$, where
$t\in (-\infty,+\infty)$, will lead to the time-independent decay rates
${\rm R}(f_{1},f_{2})_{C}$ in Eq. (7). Eqs. (8) and (9) are two basic
formulae for investigating coherent $B^{0}\bar{B}^{0}$ (or $K^{0}\bar{K}^{0}$)
decays at asymmetric $B$ (or $\phi$) factories.

	Another possibility is to measure the time-integrated decay rates
of $(\pp)_{C}$ with a proper time cut, which can sometimes increase the
sizes of $CP$ asymmetries [10].
In practice, appropriate time cuts can also suppress background and improve
statistic accuracy
of signals. If the decay events in the time region $t\in [+t_{0}, +\infty)$
or $t\in (-\infty, -t_{0}]$ are used, where $t_{0}\geq 0$,
the respective decay rates can be defined by
\be
\ba{rcl}
\hat{\rm R}(f_{1},f_{2}; + t_{0})_{C} & \equiv &
\dis\int^{+\infty}_{+t_{0}} {\rm R}(f_{1},f_{2}; t)_{C} {\rm d}t \; , \\
\hat{\rm R}(f_{1},f_{2}; - t_{0})_{C} & \equiv &
\dis\int^{-t_{0}}_{-\infty} {\rm R}(f_{1},f_{2}; t)_{C} {\rm d}t \; .
\ea
\ee
{}From Eqs. (8) and (9) we obtain
\be
\ba{rcl}
\hat{\rm R}(f_{1},f_{2}; \pm t_{0})_{-} & \propto &
|A_{f_{1}}|^{2}|A_{f_{2}}|^{2}
e^{-\G t_{0}} \l [ \dis\f{ |\x_{-} \pm \z_{-}|^{2} }{4(1+y)}e^{-y\G t_{0}}
+ \dis\f{ |\x_{-} \mp \z_{-}|^{2} }{4(1-y)}e^{+y\G t_{0}} \r . \\
& & \l .  - \dis\f{ |\x_{-}|^{2}-|\z_{-}|^{2} }{2\sqrt{1+x^{2}}}
\cos \l (x\G t_{0} +\phi_{x}\r ) \pm \dis\f{ {\rm Im}\l (\x^{*}_{-}\z_{-}\r ) }
{\sqrt{1+x^{2}}} \sin (x\G t_{0} +\phi_{x}) \r ]  \;
\ea
\ee
and
\be
\ba{rcl}
\hat{\rm R}(f_{1},f_{2}; \pm t_{0})_{+} & \propto &
|A_{f_{1}}|^{2}|A_{f_{2}}|^{2}
e^{-\G t_{0}} \l [ \dis\f{ |\x_{+}+\z_{+}|^{2} }{4(1+y)^{2}}e^{-y\G t_{0}}
+ \dis\f{ |\x_{+}-\z_{+}|^{2} }{4(1-y)^{2}}e^{+y\G t_{0}} \r . \\
& & \l .  - \dis\f{ |\x_{+}|^{2}-|\z_{+}|^{2} }{2 (1+x^{2})}
\cos \l (x\G t_{0} + 2\phi_{x}\r ) + \dis\f{ {\rm Im}\l (\x^{*}_{+}\z_{+}\r ) }
{1+x^{2}} \sin (x\G t_{0} + 2\phi_{x}) \r ]  \; .
\ea
\ee
It is easy to check that
\be
\hat{\rm R}(f_{1},f_{2};+0)_{C} + \hat{\rm R}(f_{1},f_{2};-0)_{C}
\; =\; {\rm R}(f_{1},f_{2})_{C} \; .
\ee
One can observe that in $\hat{\rm R}(f_{1},f_{2};\pm t_{0})_{C}$ different
terms are
sensitive to the time cut $t_{0}$ in different ways. Thus it is possible to
enhance a $CP$ violating term (and suppress the others) via a suitable cut
$t_{0}$ .

The formulae given above are applicable to all coherent decays of the
$K^{0}-\bar{K}^{0}$, $D^{0}-\bar{D}^{0}$, $\Ba-\Bb$, and
$B^{0}_{s}-\bar{B}^{0}_{s}$
systems. Denoting the decay amplitudes of $P_{n}\rw f_{i}$ by $A^{(n)}_{f_{i}}$
($n, i=1,2$) and the ratio of $A^{(2)}_{f_{i}}$ to $A^{(1)}_{f_{i}}$ by
$\eta^{~}_{f_{i}}$, one can also express the joint decay rates in terms of
$A^{(n)}_{f_{i}}$ and $\eta^{~}_{f_{i}}$ through the following transformations:
\be
A_{f_{i}} \; =\; \f{1}{2p}\l [A^{(1)}_{f_{i}} + A^{(2)}_{f_{i}}\r ] \; ,
\;\;\;\;\;\; \bar{A}_{f_{i}} \; =\; \f{1}{2q} \l [A^{(1)}_{f_{i}}
- A^{(2)}_{f_{i}}\r ] \; , \;\;\;\;\;\;
\rho^{~}_{f_{i}}\; =\; \f{p}{q}\f{1-\eta^{~}_{f_{i}}}{1+\eta^{~}_{f_{i}}} \; .
\ee
Such notations are usually favoured in the $K^{0}-\bar{K}^{0}$ system [6].
In the following we shall apply the above formalism
to the coherent $(\pair)_{C}$ decays and $CP$ violation at the $\U$ resonance,
a basis of the forthcoming $B$ factories [4,9]. \\

\begin{flushleft}
{\Large\bf 3. Signals of $CP$ violation at $B$ factories}
\end{flushleft}

	The unique experimental advantages of studying $b$-quark physics
at the $\Upsilon (4S)$ resonance are well known. For symmetric $e^{+}e^{-}$
collisions
the produced $B$ mesons are almost at rest and their mean
decay length is only about 20$\mu$m, a distance which is insufficient for
identifying the decay vertices or measuring the decay time difference [9,10].
If the
colliding $e^{+}$ and $e^{-}$ beams have different energies, the product of
collisions will move with a significant relativistic boost factor in the
laboratory\footnote{For a moving $\Upsilon (4S)$ system, the momentum of the
$B$ mesons in the $\Upsilon (4S)$ rest frame can be ignored. This safe
approximation
has been discussed in Refs. [9,10].}
(along the direction of the more energetic beam). This can cause the two
$B$ mesons far apart in space, such that the distance between
their decay vertices becomes measurable. It is then possible to study the time
distribution
of the joint decay rates and $CP$ asymmetries.

\begin{center}
{\large\bf A. $~$ $CP$ violation in $\Ba-\Bb$ mixing}
\end{center}

We first consider the joint decays $(\pair)_{C}$ $\rw$
$(l^{\pm}X^{\mp}_{a})(l^{\pm}X^{\mp}_{b})$,
which lead to dilepton events in the final states. Keeping the $\D Q = \D B$
rule
and $CPT$ symmetry, we have
\be
\ba{rcl}
\la l^{+}X^{-}_{i}|H|\Ba\ra & = & \la l^{-}X^{+}_{i}|H|\Bb\ra \; \equiv \;
A_{li} \; ; \\
\la l^{-}X^{+}_{i}|H|\Ba\ra & = & \la l^{+}X^{-}_{i}|H|\Bb\ra \; = \; 0 \; ,
\ea
\ee
where $i=a$ or $b$. Subsequently we use ${\rm N}^{\pm\pm}_{C}(t)$ and ${\rm
N}^{+-}_{C}(t)$ to denote the
time-dependent like-sign and opposite-sign dilepton numbers, respectvely.
Similarly,
let $\hat{\rm N}^{\pm\pm}_{C}(\pm t_{0})$ and $\hat{\rm N}^{+-}_{C}(\pm t_{0})$
denote the time-integrated
dilepton events with the time cut $t_{0}$ . With the help of Eqs. (8) and (9),
we obtain
\beq
{\rm N}^{++}_{-}(t) & \propto & |p/q |^{2} |A_{la}|^{2}|A_{lb}|^{2}e^{-\G |t|}
\l [\cosh (y\G t) -\cos (x\G t)\r ] \; , \nonumber \\
{\rm N}^{--}_{-}(t) & \propto & |q/p |^{2} |A_{la}|^{2}|A_{lb}|^{2}e^{-\G |t|}
\l [\cosh (y\G t) -\cos (x\G t)\r ] \; , \\
{\rm N}^{+-}_{-}(t) & \propto & 2 |A_{la}|^{2}|A_{lb}|^{2}e^{-\G |t|}
\l [\cosh (y\G t) +\cos (x\G t)\r ] \; ; \nonumber
\eeq
and
\beq
{\rm N}^{++}_{+}(t) & \propto & |p/q|^{2} |A_{la}|^{2}|A_{lb}|^{2}e^{-\G |t|}
\l [\dis\f{\cosh (y\G |t|) +y \sinh (y\G |t|)}{1-y^{2}}
-\dis\f{\cos (x\G |t| +\p_{x})}{\sqrt{1+x^{2}}} \r ] \; , \nonumber \\
{\rm N}^{--}_{+}(t) & \propto & |q/p|^{2} |A_{la}|^{2}|A_{lb}|^{2}e^{-\G |t|}
\l [\dis\f{\cosh (y\G |t|) +y \sinh (y\G |t|)}{1-y^{2}}
-\dis\f{\cos (x\G |t| +\p_{x})}{\sqrt{1+x^{2}}} \r ] \; , \\
{\rm N}^{+-}_{+}(t) & \propto & 2 |A_{la}|^{2}|A_{lb}|^{2}e^{-\G |t|}
\l [\dis\f{\cosh (y\G |t|) +y \sinh (y\G |t|)}{1-y^{2}}
+\dis\f{\cos (x\G |t| +\p_{x})}{\sqrt{1+x^{2}}} \r ] \; . \nonumber
\eeq
If $y$ is not very small in comparison with $x$, its size and sign should be
(in principle) determinable from the above two equations\footnote{A more
detailed
discussion has been given by Dass and Sarma in Ref. [7], where only the case of
$C$ = $-$ was taken into account.}.

Since both  ${\rm N}^{\pm\pm}_{C}(t)$ and ${\rm N}^{+-}_{C}(t)$ are even
functions of
proper time $t$, one finds
\be
\hat{\rm N}^{\pm\pm}_{C}(+t_{0})\; = \; \hat{\rm N}^{\pm\pm}_{C}(-t_{0}) \;
\equiv \;
\f{1}{2}\hat{\rm N}^{\pm\pm}_{C}(t_{0}) \; , \;\;\;\;\;\;
\hat{\rm N}^{+-}_{C}(+t_{0})\; =\; \hat{\rm N}^{+-}_{C}(-t_{0}) \; \equiv \;
\f{1}{2}\hat{\rm N}^{+-}_{C}(t_{0}) \; .
\ee
The time-integrated observables of $CP$ violation and $\Ba-\Bb$ mixing can be
defined as:
\be
\ba{rcl}
{\cal A}^{+-}_{C}(t_{0}) \; \equiv \;
\dis\f{\hat{\rm N}^{++}_{C}(t_{0}) - \hat{\rm N}^{--}_{C}(t_{0})}
{\hat{\rm N}^{++}_{C}(t_{0}) + \hat{\rm N}^{--}_{C}(t_{0})} \; , \;\;\;\;\;\;\;
{\cal S}^{+-}_{C}(t_{0}) \; \equiv \;
\dis\f{\hat{\rm N}^{++}_{C}(t_{0}) +  \hat{\rm N}^{--}_{C}(t_{0})}
{\hat{\rm N}^{+-}_{C}(t_{0})} \; .
\ea
\ee
Using Eqs. (11) and (12) we obtain
\be
{\cal A}^{+-}_{-}(t_{0}) \; =\; {\cal A}^{+-}_{+}(t_{0}) \; =\;
\f{\dis |p/q|^{2} -\dis |q/p|^{2}}
{\dis |p/q|^{2} +\dis |q/p|^{2}} \; ,
\ee
which is independent of the time cut $t_{0}$ . The nonvanishing ${\cal
A}^{+-}_{C}(t_{0})$
implies $CP$ violation in $\Ba-\Bb$ mixing. In addition, we find
\be
\ba{rcl}
{\cal S}^{+-}_{-}(t_{0}) & = & \dis\f{\dis |p/q|^{2}+\dis |q/p|^{2}}{2}
\f{\cosh (y\G t_{0}) +y\sinh (y\G t_{0}) - z \cos (x\G t_{0}+\p_{x})}
{\cosh (y\G t_{0}) +y\sinh (y\G t_{0}) + z \cos (x\G t_{0}+\p_{x})} \; , \\
{\cal S}^{+-}_{+}(t_{0}) & = & \dis\f{\dis |p/q|^{2} +\dis |q/p|^{2}}{2}
\f{(1+y^{2})\cosh (y\G t_{0}) +2y\sinh (y\G t_{0}) - z^{2} \cos (x\G
t_{0}+2\p_{x})}
{(1+y^{2})\cosh (y\G t_{0}) +2y\sinh (y\G t_{0}) + z^{2} \cos (x\G
t_{0}+2\p_{x})} \; , \\
\ea
\ee
where $z = (1-y^{2})/\sqrt{1+x^{2}}$.

In the context of the standard model, $|q/p|\approx 1$ and $y\approx 0$ are two
good approximations [3].
Thus Eq. (21) is simplified as
\be
\ba{rclcl}
{\cal S}^{+-}_{-}(t_{0}) & \approx & \dis\f{\sqrt{1+x^{2}} -\cos (x\G
t_{0}+\p_{x})}
{\sqrt{1+x^{2}} +\cos (x\G t_{0}+\p_{x})} & \stackrel{t_{0}=0}{\longrightarrow}
& \dis\f{x^{2}}{2+x^{2}} \; , \\
{\cal S}^{+-}_{+}(t_{0}) & \approx & \dis\f{(1+x^{2})-\cos (x\G t_{0}+2\p_{x})}
{(1+x^{2})+\cos (x\G t_{0}+2\p_{x})} & \stackrel{t_{0}=0}{\longrightarrow}
& \dis\f{3x^{2}+x^{4}}{2+x^{2}+x^{4}} \; .
\ea
\ee
We show the evolution of ${\cal S}^{+-}_{C}(t_{0})$ with $t_{0}$ in Fig. 1,
where the
experimental input is $x\approx 0.7$ [2]. One observes that an appropriate time
cut can significantly
increase the ratio of the same-sign dilepton events to the opposite-sign ones.
Practically time cuts should be a useful way to enhance the signals of
$\Ba-\Bb$ mixing, only if
the cost of the total number of events is not too large.

\begin{center}
{\large\bf B. $~$ $CP$ asymmetries in $\Ba$ vs $\Bb$ decays to $CP$
eigenstates}
\end{center}

Neutral $B$ decays to $CP$ eigenstates are favoured in both theory and
experiments to
study quark mixing and $CP$ violation. At the $\U$ resonance, the produced
$B^{0}_{d,{\rm phys}}$
and $\bar{B}^{0}_{d,{\rm phys}}$ mesons exist in a coherent state untill
one of them decays. Thus one can use the semileptonic decay of
one $B_{d}$ meson to tag the flavour of the other meson decaying to a
flavour-nonspecific
hadron state. Let us consider the joint transitions $(\pair)_{C}\rw
(l^{\mp}X^{\pm})f$, where
$f$ denotes a hadronic $CP$ eigenstate such as $J/\psi K_{S}, D^{+}D^{-}$,
or $\pi^{+}\pi^{-}$. To a good degree of accuracy in the standard model,
we have $|q/p|\approx 1$ and $y\approx 0$. With the help of Eqs. (8) and (9),
the time-dependent decay rates are given as
\be
{\rm R}(l^{\mp},f; t)_{-} \; \propto \;  |A_{l}|^{2}|A_{f}|^{2}
e^{-\Gamma |t|}\left [\displaystyle\f{1+|\lm|^{2}}{2} \pm
\displaystyle\f{1-|\lm|^{2}}{2} \cos (x\G t)
\mp {\rm Im}\lm \sin (x\G t) \right ]
\ee
and
\beq
{\rm R}(l^{\mp},f; t)_{+} & \propto & |A_{l}|^{2}|A_{f}|^{2}
e^{-\Gamma |t|} \left [ \displaystyle\frac{1+|\lm|^{2}}{2} \pm
\displaystyle\frac{1}
{\sqrt{1+x^{2}}}\displaystyle\frac{1-|\lm|^{2}}{2}
\cos (x\G |t| + \phi_{x}) \right .  \nonumber  \\
&  & \left . \mp \displaystyle\frac{1}{\sqrt{1+x^{2}}}{\rm Im}\lm
\sin (x\G |t| +\phi_{x})\right ] \; ,
\eeq
where $\lm = (q/p)\rho^{~}_{f}$ .
The time-dependent $CP$ asymmetries, defined by
\begin{equation}
{\cal A}_{C}(t) \; \equiv \; \frac{{\rm R}(l^{-},f; t)_{C}
- {\rm R}(l^{+},f; t)_{C}}{{\rm R}(l^{-},f; t)_{C}
+ {\rm R}(l^{+},f; t)_{C}} \; ,
\end{equation}
can be explicitly expressed as
\begin{equation}
\begin{array}{ccl}
{\cal A}_{-}(t) & = & {\cal U}_{f} \cos (x\G t) + {\cal V}_{f} \sin (x\G t) \;
,  \\
{\cal A}_{+}(t) & = & \dis\f{1}{\sqrt{1+x^{2}}}\l [ {\cal U}_{f}
\cos (x\G |t| + \phi_{x}) + {\cal V}_{f} \sin (x\G |t| + \phi_{x}) \r ] \; ,
\end{array}
\end{equation}
where
\be
{\cal U}_{f} \; =\; \f{1-|\lm|^{2}}{1+|\lm|^{2}} \; ,\;\;\;\;\;\;
{\cal V}_{f} \; =\; \f{-2{\rm Im}\lm}{1+|\lm|^{2}} \; .
\ee
We find that ${\cal A}_{C}(t)$ contains both the direct $CP$ asymmetry in the
decay
amplitude ($|\lm|\neq 1$ or ${\cal U}_{f}\neq 0$) and the indirect one from
interference of
mixing and decay (${\rm Im}\lm\neq 0$ or ${\cal V}_{f}\neq 0$). Measuring the
time distribution
of ${\cal A}_{\pm}(t)$ can distinguish between these two sources of $CP$
violation [8].

	There are two ways to combine the time-integrated decay events
(with the time cuts $\pm t_{0}$), leading to two types of $CP$ asymmetries:
\be
\ba{rcl}
{\cal A}^{(1)}_{C}(t_{0}) & \equiv & \dis\f{ \l [ \hat{\rm R}(l^{-},f;+t_{0})
+ \hat{\rm R}(l^{-},f;-t_{0})\r ] - \l [ \hat{\rm R}(l^{+},f;+t_{0})
+ \hat{\rm R}(l^{+},f;-t_{0})\r ] }
{ \l [ \hat{\rm R}(l^{-},f;+t_{0})
+ \hat{\rm R}(l^{-},f;-t_{0})\r ] + \l [ \hat{\rm R}(l^{+},f;+t_{0})
+ \hat{\rm R}(l^{+},f;-t_{0})\r ] } \; , \\
& & \\
{\cal A}^{(2)}_{C}(t_{0}) & \equiv & \dis\f{ \l [ \hat{\rm R}(l^{-},f;+t_{0})
+ \hat{\rm R}(l^{+},f;-t_{0})\r ] - \l [ \hat{\rm R}(l^{+},f;+t_{0})
+ \hat{\rm R}(l^{-},f;-t_{0})\r ] }
{ \l [ \hat{\rm R}(l^{-},f;+t_{0})
+ \hat{\rm R}(l^{+},f;-t_{0})\r ] + \l [ \hat{\rm R}(l^{+},f;+t_{0})
+ \hat{\rm R}(l^{-},f;-t_{0})\r ] } \; .
\ea
\ee
With the help of Eqs. (11) and (12), we obtain
\be
{\cal A}^{(1)}_{-}(t_{0}) \; =\; \f{\cos (x\G t_{0} +\p_{x})}{\sqrt{1+x^{2}}}
{\cal U}_{f} \; , \;\;\;\;\;\;\;
{\cal A}^{(2)}_{-}(t_{0}) \; =\; \f{\sin (x\G t_{0} +\p_{x})}{\sqrt{1+x^{2}}}
{\cal V}_{f} \; ;
\ee
and
\be
{\cal A}^{(1)}_{+}(t_{0}) \; =\; \f{\cos (x\G t_{0} +2\p_{x})}{1+x^{2}}
{\cal U}_{f} + \f{\sin (x\G t_{0} +2\p_{x})}{1+x^{2}} {\cal V}_{f} \;
,\;\;\;\;\;\;\;
{\cal A}^{(2)}_{+}(t_{0}) \; =\; 0 \; .
\ee
Clearly the asymmetries ${\cal A}^{(1)}_{-}(t_{0})$ and
${\cal A}^{(2)}_{-}(t_{0})$ signify direct and indirect $CP$ violation,
respectively. They can be separated from each other on the $\U$ resonance.
In Fig. 2 we show the ratios ${\cal A}^{(n)}_{-}(t_{0})
/{\cal A}^{(n)}_{-}(0)$ ($n=1$ and 2) as functions of $t_{0}$ . A proper time
cut
can certainly increase the $CP$ asymmetries (at some cost of decay events). In
practice, it
can also suppress background and improve statistic accuracy of signals.
Note that a suitable cut of decay time is (in principle) able to isolate the
direct or indirect
$CP$ asymmetry in ${\cal A}^{(1)}_{+}(t_{0})$. For example,
\be
{\cal A}^{(1)}_{+}\l (\f{\pi}{x\G}-\f{2\p_{x}}{x\G}\r ) \; =\;
-\f{1}{1+x^{2}}{\cal U}_{f} \; , \;\;\;\;\;\;
{\cal A}^{(1)}_{+}\l (\f{\pi}{2x\G}-\f{2\p_{x}}{x\G}\r ) \; =\;
\f{1}{1+x^{2}}{\cal V}_{f} \; .
\ee
At symmetric $B$ factories, it is possible to measure a large $CP$ asymmetry
${\cal A}^{(1)}_{+}(0)$.

\begin{center}
{\large\bf C. $~$ $CP$-forbidden transitions}
\end{center}

We finally consider the $CP$-forbidden decay modes
\be
(\pair)_{\mp} \; \rightarrow \; (f_{a}f_{b})_{\pm} \; ,
\ee
where $f_{a,b}$ denote the $CP$ eigenstates with the same or opposite $CP$
parities.
It should be emphasized that for such decays $CP$ violating signals can be
established by measuring
the joint decay rates other than the decay rate asymmetries [3]. In practice,
this implies that
neither flavour tagging nor time-dependent measurements are necessary.

	On the $\U$ resonance, the typical $CP$-forbidden channels include
$(\pair)_{-}\rw (ff)_{+}$ with
$f=J/\psi K_{S}$, $J/\psi K_{L}$, $D^{+}D^{-}$, and $\pi^{+}\pi^{-}$.
Taking $f=\Xc K_{S}$ for example, where $\Xc$ denote all the possible
charmonium states that can form the odd $CP$ eigenstates with $K_{S}$ (see Fig.
3),
we have the safe approximation $\rho^{~}_{\Xc K_{S}}\approx -1$. In addition,
we take $y\approx 0$ and $q/p \approx e^{-2i\b}$, where $\b$ corresponds to an
inner
angle of the Kobayashi-Maskawa unitarity triangle [2]. With the help of Eqs.
(7) and (8), one obtains
the branching fractions
\be
\ba{rcl}
{\rm B}(\Xc K_{S}, \Xc K_{S}; t)_{-} & \propto & {\rm B}^{2}_{\Xc K_{S}}
\sin^{2}(2\b) e^{-\G |t|} [ 1-\cos (x\G t) ] \; , \\
{\rm B}(\Xc K_{S}, \Xc K_{S})_{-} & \propto & {\rm B}^{2}_{\Xc K_{S}}
\sin^{2}(2\b) \dis\f{x^{2}}{1+x^{2}} \; ,
\ea
\ee
where ${\rm B}_{\Xc K_{S}}$ denotes the branching ratio of $\Ba\rw \Xc K_{S}$ .
Clearly the above joint decay rates are forbidden by $CP$ symmetry
($\b =0$ or $\pm\pi$). In practice, summing over the available final states
$\Xc K_{S}$ can statistically increase the number of decay events:
\be
{\rm B}_{-} \; \equiv \; \sum_{\Xc}{\rm B}(\Xc K_{S}, \Xc K_{S})_{-} \; \propto
\;
\f{x^{2}}{1+x^{2}} \sin^{2}(2\b) \sum_{\Xc}\l ({\rm B}^{2}_{\Xc K_{S}}\r ) \; .
\ee

	Just above the $\U$ resonance, an interesting type of $CP$-forbidden channels
should be $(\pair)_{+}$ $\rw$ $[(\Xc K_{S}) (\Xc K_{L})]_{-}$. Neglecting $CP$
violation
in the kaon system, we have $\rho^{~}_{\Xc K_{L}}\approx -\rho^{~}_{\Xc K_{S}}
\approx 1$ and ${\rm B}_{\Xc K_{S}}\approx {\rm B}_{\Xc K_{L}}$ to a good
degree of
accuracy. From Eqs. (7) and (9), it is staightforward to obtain
\be
\ba{rcl}
{\rm B}(\Xc K_{S}, \Xc K_{L}; t)_{+} & \propto & {\rm B}^{2}_{\Xc K_{S}}
\sin^{2}(2\b) e^{-\G |t|} \l [ 1-\dis\f{\cos (x\G |t| +\p_{x})}
{\sqrt{1+x^{2}}} \r ] \; , \\
{\rm B}(\Xc K_{S}, \Xc K_{L})_{+} & \propto & {\rm B}^{2}_{\Xc K_{S}}
\sin^{2}(2\b) \dis\f{3x^{2}+x^{4}}{1+2x^{2}+x^{4}} \; .
\ea
\ee
Summing over the possible states $(\Xc K_{S})(\Xc K_{L})$, we find
\be
{\rm B}_{+} \; \equiv \; \sum_{\Xc}{\rm B}(\Xc K_{S}, \Xc K_{L})_{+} \; \propto
\;
\f{3x^{2}+x^{4}}{1+2x^{2}+x^{4}} \sin^{2}(2\b) \sum_{\Xc}\l ({\rm B}^{2}_{\Xc
K_{S}}\r ) \; .
\ee
In Fig. 4 we show the relative sizes of the effective branching fractions ${\rm
B}_{-}$
and ${\rm B}_{+}$ in the region $0.17 \leq \sin (2\b) \leq 0.99$, limited by
the current data [11].
For our purpose, the suitable $\Xc$ states include $J/\psi$, $\psi^{'}$,
$\psi^{''}$,
$\eta_{c}$, $\eta^{'}_{c}$, etc\footnote{Note that
$\psi^{'}\rw J/\psi \pi\pi$, $\psi^{''}\rw D\bar{D}$, and $\eta^{'}_{c}\rw
\eta_{c}\pi\pi$ .}.
Since such channels occur through the same tree-level quark diagram (see Fig.
3),
their branching ratios ${\rm B}_{\Xc K_{S}}$ (or ${\rm B}_{\Xc K_{L}}$) are
expected
to be the same order. Thus a combination of the possible
decays $(\pair)_{C}\rw \Xc K_{S}$ or $\Xc K_{L}$ should increase the decay
events of
a single mode by several times. In a similar way one can study the joint
transitions
$(\pair)_{C}\rw \Xc K_{S}\pi^{0}$ or $\Xc K_{L}\pi^{0}$. For a more detailed
discussion
about $CP$ violation in the semi-inclusive decays $\Ba$ vs $\Bb\rw
(\bar{c}c)K_{S}$ or
$(\bar{c}c)K_{L}$, we refer the reader to Ref. [12]. \\

\begin{flushleft}
{\Large\bf 4. Summary}
\end{flushleft}

In keeping with the experimental efforts to study flavour mixing and $CP$
violation,
we have presented a generic formalism for the time-dependent and
time-integrated decays of
all possible $\Pa-\Pb$ systems. This formalism is useful for various
phenomenological
applications at the forthcoming flavour factories, where a large amount of
coherent $\Pa\Pb$ events will be produced and accumulated. In our calculations,
the $\D Q = \D P$ rule and $CPT$ symmetry have been assumed. Relaxing these two
constraints one can obtain the more general formulae, which should be useful
for
searching for $CPT$ violation or $\D Q =\D P$ breaking in the
$K^{0}-\bar{K}^{0}$ [6] and $B^{0}-\bar{B}^{0}$ systems [13,14].

To meet various possible measurements of $CP$ violation at asymmetric
$B$ factories, we have carried out a reanalysis of three types of $CP$
violating signals
manifesting in the coherent $\Ba\Bb$ transitions at the $\U$ resonance.
Although some comprehensive works have been done on this topic, our present
one differs from them in several aspects. We illustrate the advantage of proper
time
cuts for measuring the $\Ba-\Bb$ mixing parameters and $CP$ asymmetries.
It is shown that direct and indirect $CP$ violating effects are distinguishable
in neutral $B$ decays to $CP$ eigenstates. In addition, we explore the
possibility
to measure some $CP$-forbidden processes as a direct test of $CP$ symmetry
breaking
at $B$ factories.

A more detailed study of the topic under discussion, together with some
numerical predictions
or estimates, is in preparation. \\

\begin{flushleft}
{\Large\bf Acknowledgements}
\end{flushleft}

	I would like to thank Professor H. Fritzsch for his kind hospitality
and constant encouragements. I am also grateful to Professor D. Hitlin for
sending me some useful literature of the SLAC $B$ factory program, and to
Professors J. Bernab$\rm\acute{e}$u and D. D. Wu for beneficial communications.
I am finally indebted to the Alexander von Humboldt Foundation for its
financial support.

\newpage

\begin{figure}
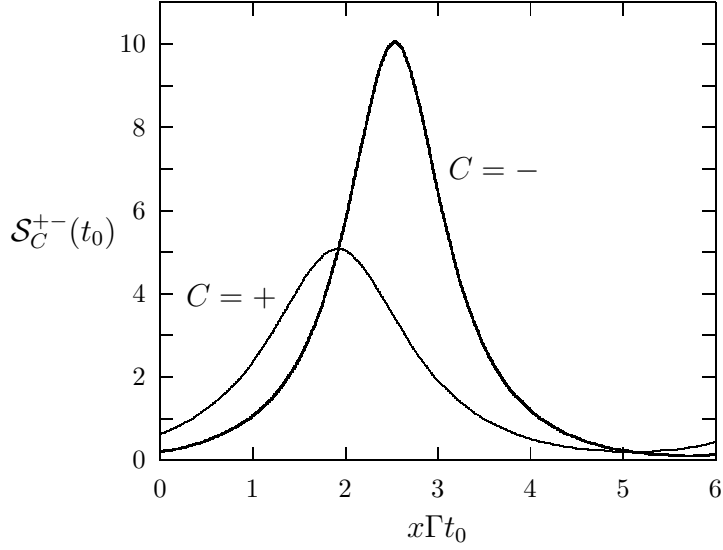

\setlength{\unitlength}{0.240900pt}
\ifx\plotpoint\undefined\newsavebox{\plotpoint}\fi
\sbox{\plotpoint}{\rule[-0.175pt]{0.350pt}{0.350pt}}%

\caption{Ratios of the same-sign dilepton events to the opposite-sign ones as
functions of the
time cut $t_{0}$ at the $\Upsilon (4S)$ resonance.}
\end{figure}

\begin{figure}
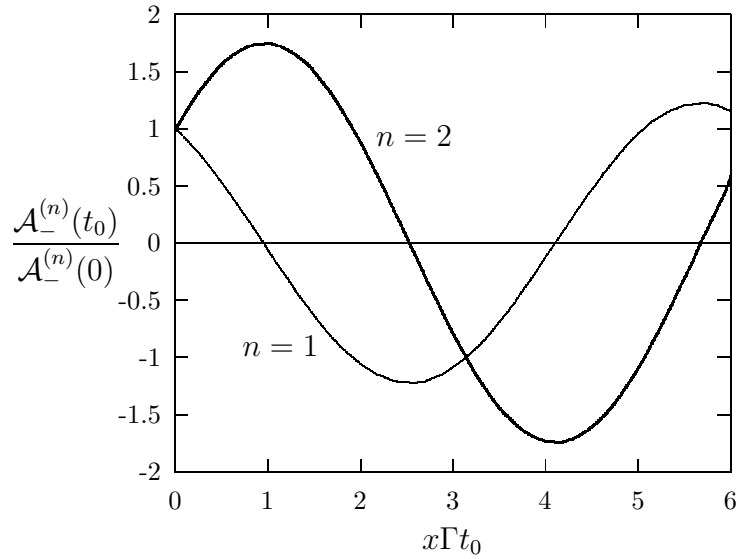

\setlength{\unitlength}{0.240900pt}
\ifx\plotpoint\undefined\newsavebox{\plotpoint}\fi
\sbox{\plotpoint}{\rule[-0.175pt]{0.350pt}{0.350pt}}%
%
\caption{Ratios of the $CP$ asymmetries ${\cal A}^{(n)}_{-}(t_{0})$
to ${\cal A}^{(n)}_{-}(0)$ as functions of the time cut $t_{0}$ on the
$\Upsilon (4S)$
resonance.}
\end{figure}

\begin{figure}
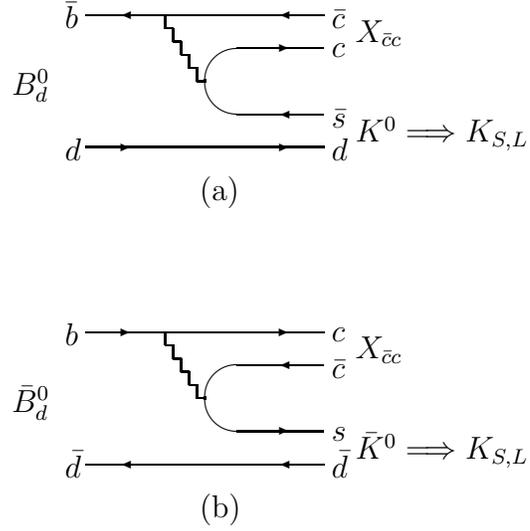

%
\vspace{-4.5cm}
\caption{ Quark diagrams for $B^{0}_{d}$ versus $\bar{B}^{0}_{d}$
decays to $X_{\bar{c}c}K_{S,L}$. Here $X_{\bar{c}c}$ denote the possible
charmonium states
that can form the odd (even) $CP$ eigenstates with $K_{S}$ ($K_{L}$).}
\end{figure}

\vspace{-3cm}

\begin{figure}
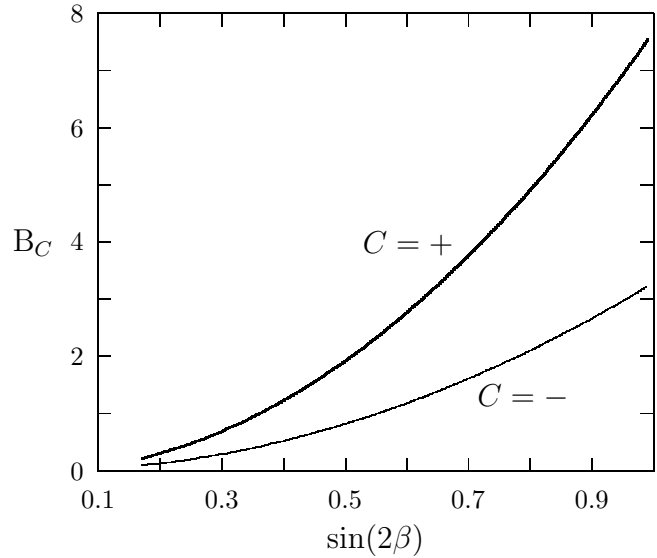

\setlength{\unitlength}{0.240900pt}
\ifx\plotpoint\undefined\newsavebox{\plotpoint}\fi
\sbox{\plotpoint}{\rule[-0.175pt]{0.350pt}{0.350pt}}%
%
\caption{Relative sizes of the effective branching fractions ${\rm B}_{C}$
(in arbitrary units) as functions of the $CP$ violating angle $\beta$
at the $\Upsilon (4S)$ resonance.}
\end{figure}

\end{document}